\documentclass[a4paper,11pt]{article}
\usepackage[utf8]{inputenc}
\usepackage[english]{babel}
\usepackage{natbib}
\usepackage{graphicx}
\usepackage{subcaption}
\usepackage{placeins}

\usepackage{wrapfig,array} %
\usepackage{amsmath} %

\title{Beauty beacon: correlated strategies for the Fisher runaway process}
 \author{ Daniil Ryabko \\{\em\small Fishlife Research } \\  {\em\small INRIA, France (on leave from)} \\ {\small daniil@ryabko.net} 
 \and Angustias Vaca \\ {\em\small Fishlife Research} \\{\small mangustias@fishlife.info}
 \and Prudencio Pazoca \\ {\em\small Fishlife Research} \\{\small ppazoca@fishlife.info}}

\date{}

\begin{document}

\maketitle
\begin{abstract}
Suppose that females  choose males based on attributes that do not signal any genetic quality that is not related to the choice itself. Can being choosy confer selective advantage in this situation?
We introduce correlated strategies, which means that females, when making their choice,  may take into consideration external and independent random factors that are known to be observable by all. Individual-based simulation is used to show that, in this case, choosiness can emerge against the cost of over 25\% when pitted against randomly mating females. Moreover,  after being established in the population, it can sustain  costs of over 35\% .
 While such costs are not biologically plausible,  they demonstrate unequivocally that sexual choice is a strong evolutionary force.  Thus, correlated strategies are shown to be an evolutionary tool that channels randomness from the environment into genetic diversity. In addition, it turns out that a higher number of attributes in the ornament makes the choice more advantageous, which may result in a runaway complexity of sexual traits. Implications for the evolution of (female) cognitive abilities and speciation are discussed. 
\end{abstract}


\section{Introduction}

Elaborate male ornaments and displays have fascinated and puzzled  researchers  since  \cite{darwin1871descent} proposed sexual selection as one of the main drivers of evolution.  \cite{fisher1915evolution} famously suggested an explanation that came to be known as the (Fisher) runaway process or the ``sexy sons'' hypothesis: if females choose males with some particular trait, then a genetic link is established between this choice and the trait. The choice creates a positive selection pressure on the trait, which, in turn, promotes choosiness. In other words, sons of choosy females are more likely to be chosen, as they exhibit the trait.
 Subsequently developed theoretical models \citep{o1967general,lande1981models,kirkpatrick1982sexual,bulmer1989structural} confirm this hypothesis, but also show that the runaway process quickly reaches a point when the choice is no longer advantageous: either all males already have the trait, or a balance is reached where the (possible) cost of the trait compensate the selective advantage it provides (see also \citealp{kuijper2012guide} for a review). 
Thus,  if the choice is costly, then at equilibrium it is eliminated~-- a problem that came to be known as the lek paradox  \citep{kirkpatrick1991evolution}.  A solution has been proposed by  \cite{pomiankowski1991evolution,pomiankowski1995resolution} in the form of biased mutation (against the chosen trait), though the costs it allows for are very small if realistic values of mutation bias are considered (e.g., \citealp{kokko2015mate}).
Subsequent approaches developed in the literature include   dynamic costs of choice \citep{kokko2015mate} and   frequency-dependent choice strategies, in particular, selecting for the rarer male type \citep{kokko2007evolution}.  There is, however, still no consensus among evolutionists on the question: can choosiness  be advantageous if the traits on which the choice is based  do not confer any other information except related to the choice itself (the pure ``sexy sons'' explanation of the Fisher runaway)?  Considering  the ``no'' camp of this question,  the ``good genes'' model posits that the sexual selection is for traits that somehow signal higher fitness. The ``genetic capture'' hypothesis proposes a resolution to the lek paradox within this model, suggesting that male sexual traits depend on the overall conditions and thus on many loci, thereby maintaining genetic variation by mutation \citep{rowe1996lek,tomkins2004genic}. Another related approach is ``genotype-by-environment'' interaction \citep{danielson2006genotype}, which suggests some genotypes are better in some environments and others in others, so that genetic variation is maintained in heterogenic or changing environments; \cite{kokko2008condition} attempt to unify the two latter models. 
A recent attempt at combining the ``sexy sons'' and ``good genes'' approaches was made by  \cite{henshaw2022evolution}, who uses individual-based simulation to 
conclude that arbitrary ornaments are insufficient to sustain highly costly mate search, but can ``piggy back'' on  quality-dependent ornaments.

Key to all these approaches %
is the question of how genetic variation is maintained. The Fisher runaway process needs somewhere to run, the choosy females need something to select from~--- the male population should not be uniform.  Moreover, if we step back from the lek paradox and consider the original phenomenon, the puzzling part is not that the female choice persists, but the  sheer complexity of male ornaments and displays. It is particularly puzzling since often closely related species differ greatly in male sexual traits \citep{hulse2020sexual}. The question of how genetic variation is maintained in natural populations is sometimes called  one of the most fundamental evolutionary questions  \citep{danielson2006genotype}. Choice depletes genetic variation, but, at the same time, the choice depends on the variation to exist.
Despite a variety of mechanisms that have been proposed to address it (e.g., \citealp{radwan2008maintenance}), this 
problem still appears unresolved \citep{barton1989evolutionary,barton2002understanding,kaufmann2023sexually}.

Thus, the {\em main problems addressed} in this work are the maintenance of genetic variation in populations subjected to sexual selection, and, more specifically, the lek paradox.

 To understand the suggested solution, first  consider the following purely imaginary situation. Males and females congregate on a lek. Males have observable ornaments, and  females are to select with whom to mate. Every female breeds once per season.   Before the mating season  starts, a Genie  visits the lek and  pins a highly visible poster  to a tree. The image on the poster is that of a male with a particular ornament, which we shall call the {\em beauty beacon}.  Females may ignore this image altogether and mate completely at random; for the sake of this example, we  can assume that they may also make their choice based on any private criterion they wish to employ. However, they may also decide to base their choice solely on the beauty beacon: choose the male that most resembles the image displayed;  call this {\em the beacon strategy}. Moreover, the Genie keeps changing the beauty beacon every few seasons, either entirely, or by altering some detail of the male's ornament on the poster image at random. The Genie is, thus, a fashion dictator of sorts~--- but, of course, relevant only to those who choose to follow fashion.
There is yet more to the story, as the Genie  is evil: each season, he shall kill each female that uses the beacon strategy with a probability of, say, $c$ (drawn independently for each female).
 It may seem that there is no reason for the females to follow the beacon strategy if $c$ is anything but 0 (and why even then?). In particular, every time the Genie changes the beacon, the ``sexy sons'' link between the ornament and choosiness is  weakened (if the change is small) or broken (if it is complete).  Yet, the beacon strategy shall emerge, spread over the majority of the population and persist, with the values of $c$ over 25\%; and if the Genie is lenient in his killings and spares  the first several generations, allowing the beacon strategy to emerge and spread, then this strategy persists with $c$ even higher: well over 35\% (as shown in the experiments below, even over 40\% for some parameter values).

  The intuitive explanation why this works is that the changing fashion, here dictated by the Genie and his poster beacon, by moving the target of the Fisher runaway process makes sure that there is always somewhere to run; in other words,  the Genie helps the population maintain its genetic diversity by providing a common source of randomness. Comparing it to the 
   biased-mutation solution to the lek paradox   \citep{pomiankowski1991evolution}, the latter makes sure that there is still some genetic diversity closer to the end of the runaway process,  that is to say, when almost all but not all males have the selected trait. However, the end of the process is not where the selection pressure is the strongest, and so we shall see that the cost of choosiness that can be sustained is much lower. By varying the parameters of the way the Genie dictates the fashion and changes the beacon, he may sustain the genetic diversity and thus the runaway process closer to the maximum of the selection pressure.

  Correlated strategies in game theory were introduced by  \cite{aumann1974subjectivity}. A correlated strategy allows the participants to a game to use their observations of a public random signal. In general, the players may have different private observations of this signal, but for our purposes a fully observable public random event is sufficient.  Simply put,  the players may agree to flip a coin. A strategy based on this coin flip may be an equilibrium strategy, that is, each player stands to lose by deviating from it; the resulting notion of correlated equilibrium generalizes that of Nash equilibrium. The following example from \citep{gintis2000game} illustrates the concept. 
There are two players, A and B; A can use the strategies up or down, and B can use the strategies left or right. The matrix of payouts is as follows. 
\begin{wrapfigure}{l}{5cm} %
\centering
\begin{tabular}{r|c|c|}
 & $l$ & $r$ \\\hline
$u$ & (5,1) & (0,0)  \\
$d$  & (4,4) & (1,5)
\end{tabular}
\end{wrapfigure}
 First, consider the strategies without the use of common randomness. Both (5,1) and (1,5) are pure-strategy  equilibria, and (2.5,2.5) is the payout of the mixed-stretegy equilibrium where each player plays each strategy with  probability 1/2 independently of each other. Now, if the two players can both observe a single unbiased coin-flip, and play $(u,l)$ on {\em heads} and $(d,r)$ on {\em tails}, then the expected payout becomes (3,3). This is the correlated equilibrium of the game.

 In the context of evolutionary strategies, the players that use the same strategy do so simply because they have the same genes. (Different genes mean possibly different strategies.) Thus, arbitrary complex strategies may develop and no agreements between the participants are necessary for them to use the same strategy. Next up is the  question of where to find  the Genie and his poster tree. The answer is that no explicit image of the ideal male is necessary. What is sufficient is simply a commonly observable source of randomness. The female strategy is then any function that maps this source of randomness into male traits.  Random, or (equivalently, for our purposes) unpredictable  observations are abound in nature: they may be related to the weather (temperature, pressure, humidity, winds, etc.), the condition of the populations of other species (including vegetation)  or any other changing  aspect of the environment. Any such aspect may be used, and the strategy is simply a mapping from this source of randomness into male features. What is important is that the random source changes with sufficient but not excessive frequency. We shall see from the model what this means. 
 There is a growing body of evidence in the literature that both male sexual traits and female preferences depend on the environment, most notably in fish \citep{seehausen2008speciation, hulse2020sexual,heuschele2009environment,justin2000communication}, but also in other animals including frogs, lizards and birds \citep{ryan1990sensory,fleishman1992influence,hernandez2021colourful}.
 We do not attempt to find any actual mappings of the environment into sexual traits; they would  be specific to both the species and the environment, but, as far as the considered sexual choice strategies are concerned,  may otherwise be arbitrary as long as they preserve the necessary randomness. 

Thus, in the experiments, we consider random source which already has the needed form of a male ornament, both represented simply by a sufficiently large array of binary loci.

However, to illustrate how a natural source of randomness can be harnessed by  a beacon strategy, we  construct an example based on the randomness present in the population itself.  
  The construction is as follows. A source of  randomness that we know to be present in every biological population is mutation. While each mutation is private, the average of each expressed feature over the population is somewhat random  and publicly observable. Thus, consider a visible feature on which there is no sexual or natural selection, and consider its {\em average} value over the population. This value is driven by the drift alone. This is a slow process, so we need  a large ensemble of such features. This ensemble is our source of randomness.
  To take a concrete, event if grotesque in its artificiality,  example, consider a peacock population in which females judge males based on the left half of their tail only. The males are never judged based on the right half of their tail; rather, the right half is used, by each female, to take an average over the whole population, or else at least over such part of the population that she can observe. This is the female-estimated beacon. She then compares the left half of the tail of each male that presents himself as a potential mate to this average over the right half that she constructed.

  In the last example,  a new challenge for the beacon strategy presents itself: the public information may actually be slightly different for different females, as each one may practically observe only a part of the male population. The same problem is undoubtedly present if other sources of public randomness are used.  We shall see from the results  that this, indeed, affects the performance of the strategy. This difficulty also sheds new light on the nature of the cost of choosiness itself, and gives rise to other strategies such as mate-choice copying.  These implications are discussed further in the last section.

The correlated female choice strategies proposed are pitted against the random-choice strategy with zero cost in an individual-based simulation, which is one of the standard tools in theoretical research on evolutionary models (see, e.g., \citealp{kuijper2012guide} for a review).  The results are unequivocal in that the choice emerges and persists despite very high costs: between 25\% and 45\% depending on the parameters of the strategies, and whether an external beacon or female-estimated beacon is used.  

It also becomes clear from the experiments that the more attributes there are to choose from, the better. Furthermore, 
for the female-estimated beacon, the mutation rate of the beacon ornament emerges as a crucial parameter. Specifically, the strategy performs well if the mutation rate is 0.01 at each locus, and rapidly deteriorates if it is decreased. While such (and higher) mutation rates are often used in theoretical models (e.g., \citealp{fawcett2011sex}), it is perhaps too high to be realistic. However,  higher mutation rate can be achieved  by increasing the number of loci on which mutation operates and combining them. Mathematically, it is easy to do: if we have $k$ binary loci with mutation rate $m$, then taking a XOR operation over them we obtain a single binary locus with a mutation rate of almost  $km$ for small $m$ and $k$. Perhaps, say, a female cichlid fish does not perform this exact arithmetic trick when choosing her mating partner, and the one she uses might be  different from those used by an anoline lizard; yet,  an operation of this complexity appears to be well within their neurological possibilities. It is worth noting that increasing the number of loci in sexual traits (increasing the ornament size) has been proposed as a part of the solution to the lek paradox already by \cite{pomiankowski1991evolution}, in particular, to explain the biased mutation rate; though it was criticized on the basis that trait expression at any given level of condition is likely to be subject to stabilizing, rather than escalating, non-linear selection \citep{rowe1996lek,tomkins2004genic}. 

Here we obtain an explanation of the large number of observable parameters in the male ornaments and displays, by providing two ways the choosy females benefit from it. First, the more attributes to base the choice on, the more advantageous is the beacon strategy. Second, mutation on the traits on which the choice is not based directly provides the needed external source of randomness on which to base the choice, again making the beacon strategy more advantageous. 
This addresses another  long-standing puzzle: the multitude of signals used for sexual choice. Courtship displays are extremely diverse, and most  occur across at least two sensory modalities, such as visual and auditory or vibratory and olfactory \citep{mitoyen2019evolution}. For example, in birds these may include not only the colorful plumage itself but also  coordinated  wing and tail movements together with  vocalizations \citep{bradbury1998principles}.
   Given the (potential) costs of producing and receiving signals, why use more than a single cue \citep{bro2010dynamics}? The answer we obtain is that these signals may be used at once as a source of randomness (the more randomness the better) and as a receiving field of genetic variation whereto this randomness is channeled. 

Finally, this highlights the complexity of the female choice itself, giving  rise to a number of intriguing questions, concerning alternative mating strategies such as mate-choice copying, and the effects of the evolution of the selection rules themselves, which may include speciation. These are further discussed in the last section.

\section{The model}\label{sec:model}

An individual-based simulation model is used to demonstrate the viability of costly beacon strategies. The parameters of the model are summarized in Table~\ref{table:params}.

The model consists of a population of a constant size $S$   with a fixed 1:1 sex ratio, and discrete generations. 
Individuals are characterized by their alleles at a number of haploid loci. These are: an array  of size $A$ (mostly 100 in the experiments) of binary  attribute loci, which represent the {\em  ornament}. These loci are only expressed in the male. In addition, there is a second array of binary loci, of the same size $A$, which is also only expressed in the male (to be used by the females with the female-estimated  beacon strategy as described below). We call it the {\em beacon ornament}.  The strategy locus, which is only expressed in the female, also has two alleles,  one representing the  random choice strategy and the other 
the  beacon strategy.  %

Each generation consists of  reproduction,  mutation and decimation  (the latter often called viability selection in the literature).

The {\em reproduction} takes place by selecting randomly a female among those currently alive, who then breeds and produces two offspring, one male and one female; the next female is selected and the  process is repeated until $S$ offspring are produced (i.e., the next generation is filled).
The  {\em breeding} follows the so-called \citep{janetos1980strategies} {\em best-of-$N$} model, %
 that is, each female is presented with a pool of $N$ randomly selected males to make a choice from.  The female chooses according to her strategy (the allele at the corresponding locus), as described below. Each offspring inherits the strategy  from one of the parents with equal probability.
 The same goes for the the ornament and the beacon ornament, but each of these are inherited in its entirety (all $A$ loci) from one of the parents.
 
{\em Mutation} in the offspring is performed independently on each locus, with probabilities $m$ for the ornament and the beacon ornament, and with probability $m_s$ for the strategy.  The {\em decimation} takes place both on females and males: each female is taken out of the reproductive pool (killed) with a certain probability $c$ which is the {\em cost of her strategy}.  The cost of the random strategy is always set to 0, while the cost of the beacon strategy is some value $c\in(0,1)$.
 For the sake of compatibility with previous models in the literature, there is a small decimation on the males, with the probability $c_m$ of being killed proportional to the number of $1$s in the ornament (thus slightly pushing towards the all-zeros ornament). 

The female strategies are as follows.  The random strategy is exactly as the name suggests. 
For the beacon strategy, we consider two variants. 
The first is called {\em global beacon}  strategy. For this strategy, all females observe the same array of binary variables of size $A$ (the same size  as the male ornament), which is the global beacon. The female using this strategy compares the ornament of each male in her pool (of $N$ randomly selected males) to the global beacon, and selects the one  closest to the global beacon. The distance is simply the average  absolute distance between the elements of the two arrays  ($l_1$ distance).
 The global beacon changes from time to time. We consider two variants of these changes: either it changes completely every $t_{change}$ generations (for example, $t_{change}=50$) to a new random array; or only a few (1-5) elements of the array are changed at random. These changes are independent of everything else.

The other variant of the beacon strategy we consider is the {\em female-estimated beacon}: each female is given a random pool of $B$ males to construct an estimate (the estimation pool).   The estimate is simply the average beacon ornament of these males. She then uses this estimated beacon in the same way the global beacon is used in the previous strategy, that is, the ornament of each of the $N$ males in her breeding pool is compared to the estimate and the closest one is selected. Note that the two pools of males available to each female~--- the breeding pool of size $N$ and the estimation pool of size $B$~--- are selected independently. There is no direct selection on the beacon ornament of the males: these are only used for estimation. 

\subsection{Estimating viability}

To evaluate the viability of the beacon strategies considered, each of the tested strategies is paired against the random-choice strategy. 
This means that the strategy locus has only two  alleles in each experiment: one corresponding to the random-choice strategy, and the other to the (beacon) strategy being tested.
The random-choice strategy  has cost 0, while the tested strategy has cost $c$. 
The {\em expected viability at cost $c$} of the tested strategy after $T$ generations with initial population $I$  is the expected ratio of the females using the tested strategy after $T$ generations starting with population $I$. Furthermore, we can define the {\em $\alpha$-critical viability} of a strategy $S$ as the largest cost $c$ with which  the  expected viability of the strategy is greater than than $\alpha$  (this variable also depends on the initial population $I$ and the number of generations $T$).

 In the experiments we consider three ways of setting the initial population: either all 0s, which means that all the ornament attributes are set to 0 and the only strategy of females is the random one (the beacon strategy tested should then emerge due to mutation); the random initialization, where every attribute is set uniformly at random (including the strategy); and the {\em deferred decimation} initialization, where the initial population is random but the cost $c$ is set to zero for the first several generations (50 in the experiments), allowing the tested strategy to become established in the population. Note that the latter initialization is similar to fixating the tested strategy at the beginning, but addresses the issue of how to  initialize the male ornaments.  As would be expected, the  viability depends on the initialization, with the zero initialization giving the lowest resulting value and deferred decimation the highest.  

Note that the expected ratio of females with the tested strategy in the definition of viability can hide extinction with non-zero probability.  Indeed, this is what happens in the experiments with some values of the parameters: the beacon strategy either dominates the population or goes to extinction. 
An alternative definition would be to consider viability at threshold $\alpha$ with confidence $\epsilon$: say that a strategy has viability $c$ at threshold    $s$ with confidence $\epsilon$ after $T$ generations and initial population $I$ if the ratio of females with this strategy with cost $c$ is at least $s$ with probability at least $1-\epsilon$ after $T$ generations when starting with initial population $I$. This value is perhaps more interesting, but since it involves yet another  parameter (or two, if we consider the threshold $\alpha$), we use the  expected viability instead.

\begin{table}\caption{Parameters of the model} \tiny
\begin{tabular}{p{10mm}|p{6cm}|p{4cm}} \label{table:params}
{ Notation } & { Parameter }& { Value(s) used in the experiments; {\bf bold} for the main one}\\\hline
$S$ & Population size &  2000, constant 1:1 sex ratio \\\hline
$A$& Number of attributes in the ornament & {\bf 100}, 1000; {\bf 1} for the biased mutation comparison \\\hline
$c$ & Cost (mortality rate) of the choosy strategy & variable \\\hline
$c_m$ & Cost (mortality rate) of males & 0.01$\times$ average ornament attribute  \\\hline
$N$ & Batch size: number of males a female can choose from & {\bf 20} \\\hline
 $m$ & Attribute mutation rate, ornament and beacon ornament & {\bf 0.01}, 0.001, 0.0001\\\hline
 $m_s$ & Strategy mutation rate & {\bf 0.01} \\\hline
$T$ & Total number of generations & 5000 \\\hline
\end{tabular}
\\ \\ \centerline{   Parameters of the beacon strategy: global beacon} \\ \\ \\ 
\begin{tabular}{p{10mm}|p{6cm}|p{4cm}}
 & Number of generations between changes & {\bf 1} (gradual change), {\bf 50} (complete change) \\ 
 & Number of attributes to change & {\bf 1} , 2, 3, 5,(gradual change)  $A$ {\bf =all} (complete change)  \\ 
\hline 
\end{tabular}
\\ \\ \centerline{   Parameters of the beacon strategy: female-estimated beacon} \\ \\ 
\begin{tabular}{p{10mm}|p{6cm}|p{4cm}}
$A$ & Number of attributes in the male's beacon& Same as the number of attributes ($A$)\\
$m$ & Beacon attribute mutation rate &Same as attribute mutation rate\\ 
$B$ & Estimation batch size: number of males used by each female to estimate the beacon& 100\\
\hline 
\end{tabular}
\\ \\ \centerline{   Parameters of the comparison strategy: biased mutation} \\ \\ \\ 
\begin{tabular}{p{10mm}|p{6cm}|p{4cm}}
 $m_{extra}$ & Probability of additional mutation 1$\to$0 & 0.45 \\ 
\hline 
\end{tabular}

\end{table}

Table~\ref{table:params} shows all the parameters of the model and their values used in the experiments. The values in bold are those used in the main series of experiments, while the rest are the values used in the experiments designed to elucidate the role of each parameter.

\subsection{Comparison model: biased mutation}
Estimating viability as described already means comparing the tested strategy to the random choice. However, it is interesting to have another comparison strategy, tested against random choice in the same manner, using the same population parameters. 

For this, we have chosen the {\em biased mutation} model: the classical model of  \cite{pomiankowski1991evolution}, in which the simple Fisher runaway process is endowed with an asymmetric mutation, biased against the ornament selected. Thus, for this model we consider a single-locus ornament ($A=1$). The rest of the parameters  are the same as in the experiments with the beacon strategies, except for an additional  mutation parameter:  the mutation bias $m_{extra}$, whose meaning is as follows. 
After the mutation on the ornament allele takes place, an additional mutation is applied. If the value of the allele is 1 (ornamented male), the allele is flipped to 0 
  with probability $m_{extra}$. 
  The female strategy here is to always prefer the ornamented male, that is, the one with allele 1. 
  
 Separate experiments were run to determine the approximate value of mutation bias $m_{extra}$  that would result in the highest  critical variability. The value found was $0.45$.   While not  biologically plausible, this value has been chosen in the experiments as the hardest to compete against for the beacon strategies.

\section{Results}

The results of the simulations show that the critical viability of the beacon strategy is between 25\% and 40\%, depending on the strategy used, as well as on the initialization as described above. The strategies that use an external  target (global beacon) have the highest viability; the viability decreases if the females have to rely on their own estimates of the beacon.  
When compared to the runaway with biased mutation, all the beacon strategies evaluated show a significantly higher viability.

\begin{figure}[h]
    \caption{\small Viability of the beacon strategy with a global beacon. The values of the  parameters are 
    $T= 5000$, $A=100$, $N=20$,  $m=m_s=0.01$, $S=2000$. Colors correspond to different initializations: red to zero initialization, green to random initialization and blue to  random initialization with  decimation deferred till generation 50.
     Confidence intervals are at 1\% with either 10 or 50 replica runs. } \label{fig:plotsGB}
\centering
    \begin{subfigure}[b]{0.5\textwidth}
  \centering
  \caption{\tiny Global beacon changes  to a completely new random array every 50 generations}
      \label{fig:plotGB2CH50}
\centering
\includegraphics[width=1\textwidth]{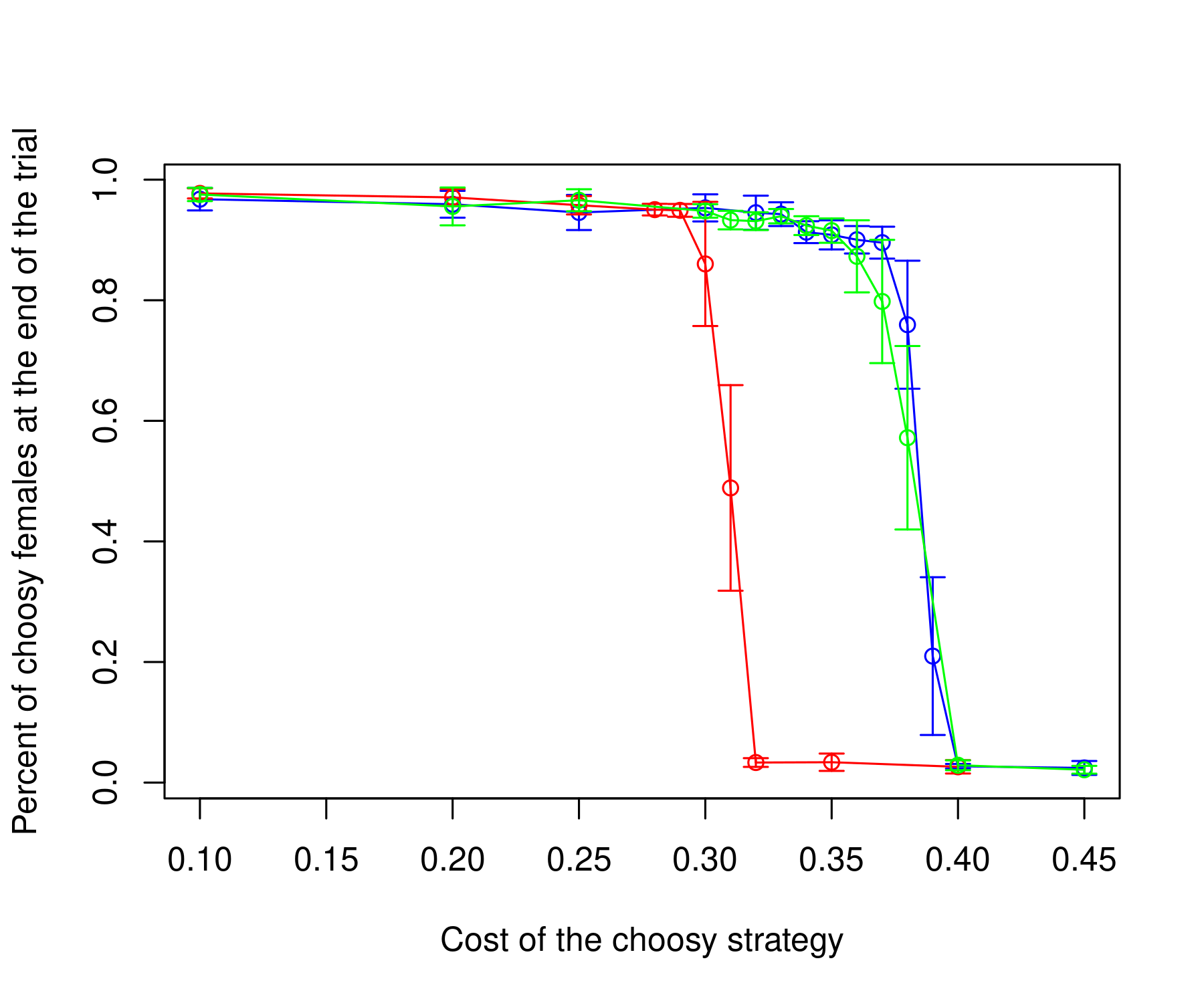}
    \end{subfigure}~
    \begin{subfigure}[b]{0.5\textwidth}
\caption{\tiny Global beacon changes every generation, one locus at a time.}    \label{fig:plotGB1Ch1}
\centering
\includegraphics[width=1\textwidth]{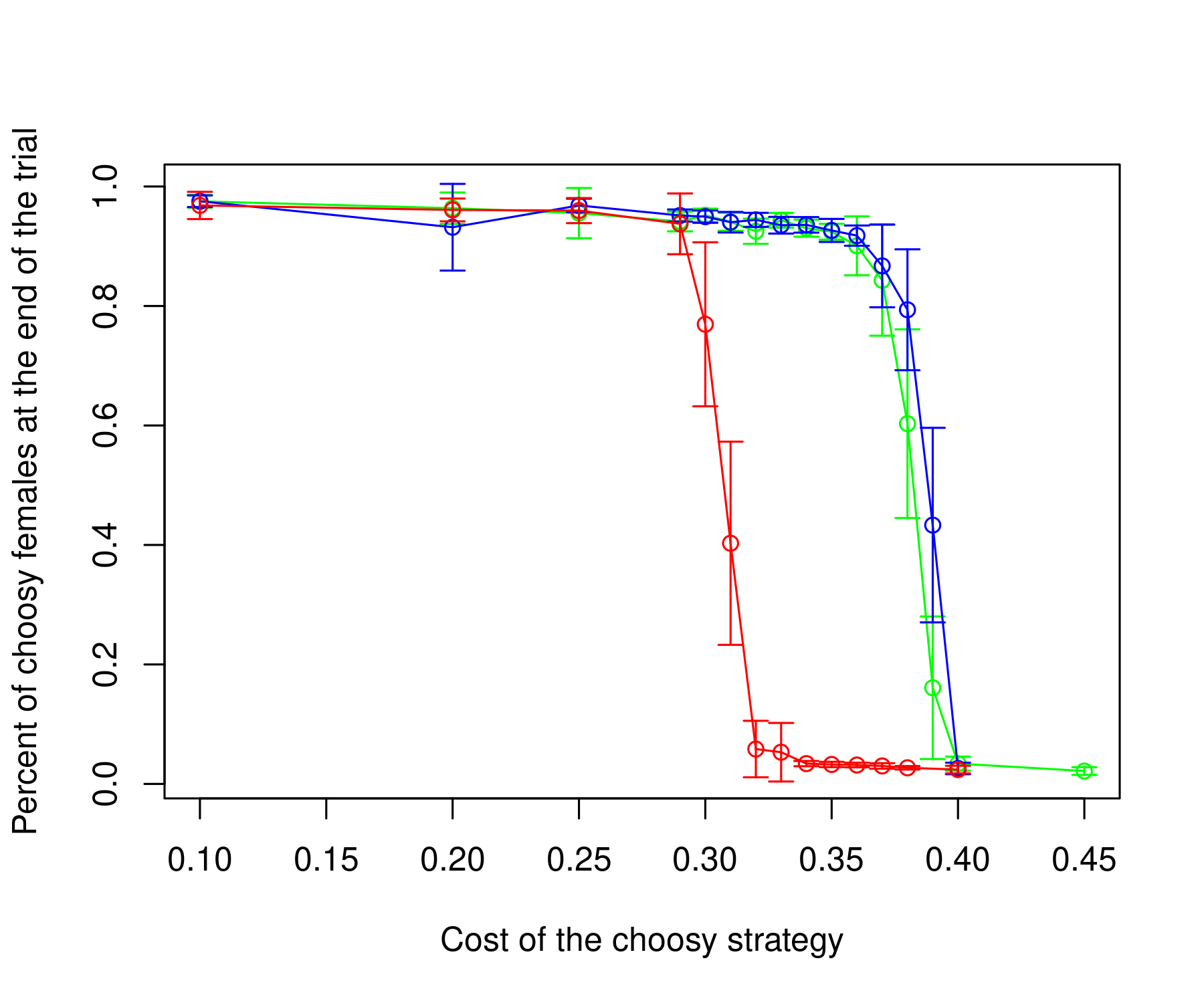}
    \end{subfigure}
\end{figure}

Figure~\ref{fig:plotsGB}  shows the viability of the beacon strategy with the global beacon. Each point corresponds to the percent of females with the beacon strategy at the end of the trial ($T$=5000 generations) for the corresponding cost $c$ of the strategy. On Figure~\ref{fig:plotGB2CH50}, the global beacon changes every 50 generations to a new, completely random one; whereas, on Figure~\ref{fig:plotGB1Ch1}, every generation one randomly selected attribute of the beacon is changed to a random value. Different colors correspond to different initialization: red to zero initialization (all females use the random choice strategy and all male ornaments are uniformly 0), green to random initialization (all values are uniformly random) and blue to  random initialization with  decimation deferred till generation 50 (the cost of the beacon strategy is 0 before that point). 
We can see that, on either of the plots, the estimated expected viability of the  beacon strategy is significantly above 70\% with the costs as high as 37\% provided it is sufficiently established in the population; if, initially, there are no females with the beacon strategy, then it emerges and persists despite the cost of up to 30\%. The difference between random initialization and deferred decimation is small, as is the difference between the two modes of changing the beacon.

\begin{figure}[h!]
\caption{\small Viability of the beacon strategy with female-estimated beacon, estimation batch size $B=100$ (solid lines). The number of loci in the ornament and  in the beacon ornament is $A=100$. Comparison model (dotted lines) is the runaway with biased mutation; here $A=1$, extra mutation bias $0.45$. 
The rest of the parameters are the same in both models:
    $T= 5000$, $N=20$,  $m=m_s=0.01$, $S=2000$. Colors correspond to different initializations: red to zero initialization, green to random initialization and blue to  random initialization with  decimation deferred till generation 50. Confidence intervals are at 1\% with either 10 or 50 replica runs.}
\label{fig:plotLocalFBwithBL}
\centering
\includegraphics[width=0.8\textwidth]{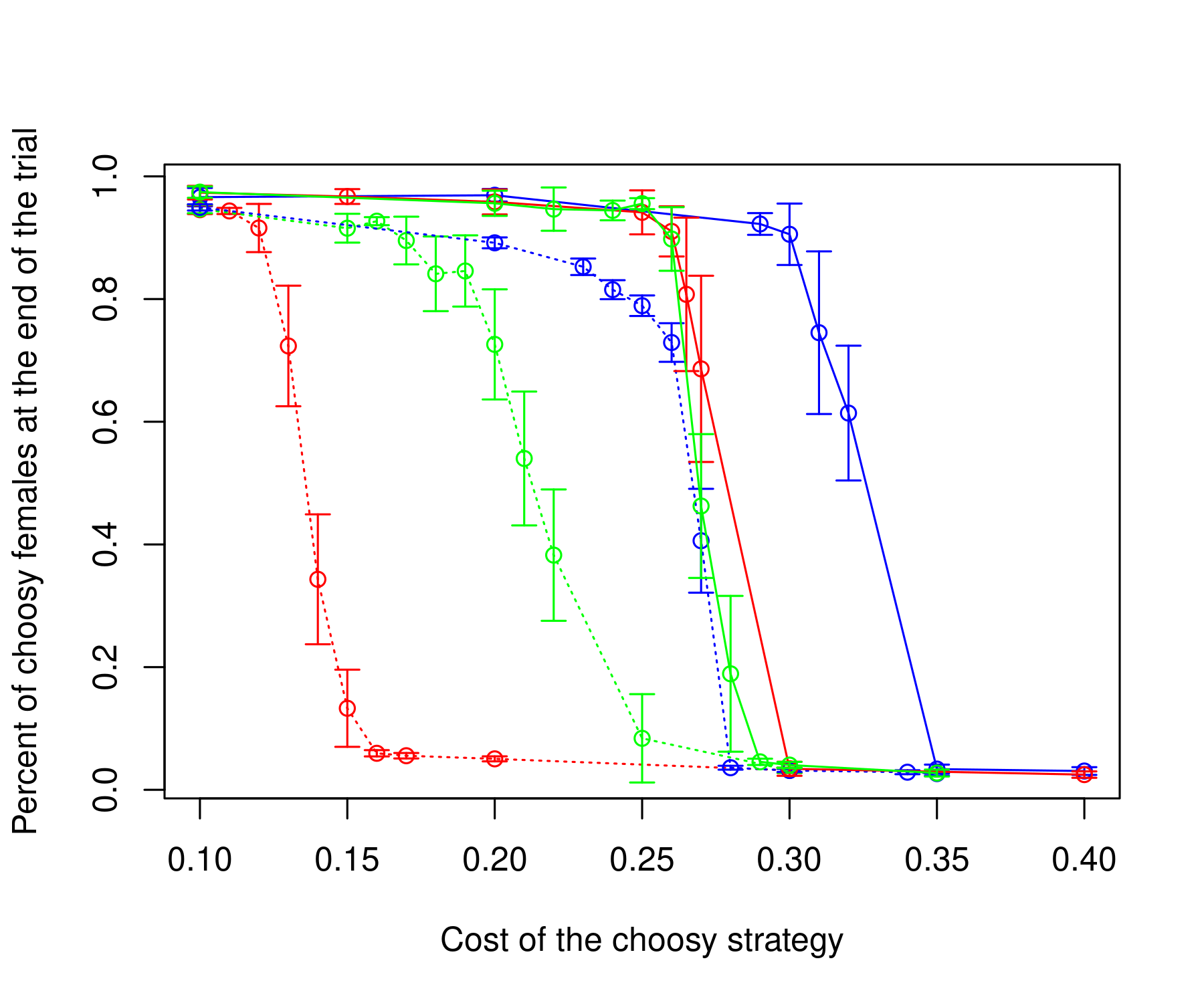}
\end{figure}

Figure~\ref{fig:plotLocalFBwithBL} shows the viability of the beacon strategy with female-estimated beacon. Thus, females using this strategy have to build their own estimate of the beacon using the beacon ornament of the males. Each one uses a separate pool of $B=100$ males to take the average of their beacon ornaments and use this as the target. The viability is plotted against the comparison model of the runaway with biased mutation. Here each male has $A=1$ ornament locus, and the females that do not choose randomly prefer males with allele 1 at this locus. The additional mutation $1\to0$ on this locus is $m_{extra}=0.45$. 
We can see that the viability of the beacon strategy is significantly higher than that of the comparison model, although if the comparison model is run with deferred decimation and the beacon strategy is run with random or zero initialization then their viability is similar. Comparing to the beacon strategy with a global beacon, we can see that the viability deteriorates if females have to use their own estimates of the beacon in the way considered here. Still, the beacon strategy with female-estimated beacon emerges and persists with costs significantly above 25\%, and, if it is sufficiently established in the population, it can  persist with cost over  30\%.

\FloatBarrier

\subsection{The role of the parameters}\label{sec:params}
The female choice strategies considered depend on a number of parameters. It is, of course, impossible  to explore exhaustively this parameter space using the experimental technique adopted. In our view, this does not diminish the value of the results, because showing that a strategy is viable with some values of the parameters demonstrates that it is evolutionary plausible~-- provided the values of the parameters are not outside the range of what is naturally possible.
This said, it is still interesting to explore the role of each of the parameters, to see what could be the evolutionary path towards the strategy studied, as well as to understand where the process may be headed. 
We attempt to do so  by varying the parameters one by one and looking how it affects the viability of the strategies.

\subsubsection{The batch size $N$ and the estimation batch size $B$}
First we take a look the role the batch size $N$ on the viability of the beacon strategies. 
Not surprisingly, the results depend strongly on this parameter, with higher batch sizes resulting in a higher viability of all beacon strategies.
This is in line with previous findings that examined the role of various parameters in  individual-based simulations \citep{roff2014evolution}.

\begin{wrapfigure}{l}{45mm}\caption{\small 0.5-critical viability of beacon strategies for various values of batch size $N$}\label{tN} %
\centering
\begin{tabular}{r|c|c}
$N$ & {\small global}  & {\tiny female-estimaed} \\\hline
2 & 0.06 & 0.04   \\
5  & 0.2 &  0.14 \\
20  & 0.37 & 0.27 \\
100  & 0.46 & 0.36 
\end{tabular}
\end{wrapfigure}

\begin{figure}
    \caption{\small Viability of the beacon strategy for different values of $N$, using a global beacon (Figures \ref{fig:plotGBLBS5},\ref{fig:plotGBLBS5},\ref{fig:plotGBLBS100}) and female-estimated beacon (Figures \ref{fig:plotFLBBS2},\ref{fig:plotFLBBS5},\ref{fig:plotFLBBS100})
    with values $N=2,5$ and $100$. The values for  $N=20$ can be seen on Figures~\ref{fig:plotGB1Ch1} and~\ref{fig:plotLocalFBwithBL}.  The values of the rest of the  parameters are     $T= 5000$, $A=100$,  $m=0.01$, $S=2000$, $B=100$ for female-etimated beacon and for the global beacon the change is every generation 3 loci at a time.    Confidence intervals are at 1\% with either 10 or 50 replica runs. } \label{fig:plotsBS}
\centering
    \begin{subfigure}[b]{0.3\textwidth}
  \centering
  \caption{\tiny Global beacon, $N=2$.}
      \label{fig:plotGBLBS2}
\centering
\includegraphics[width=1\textwidth]{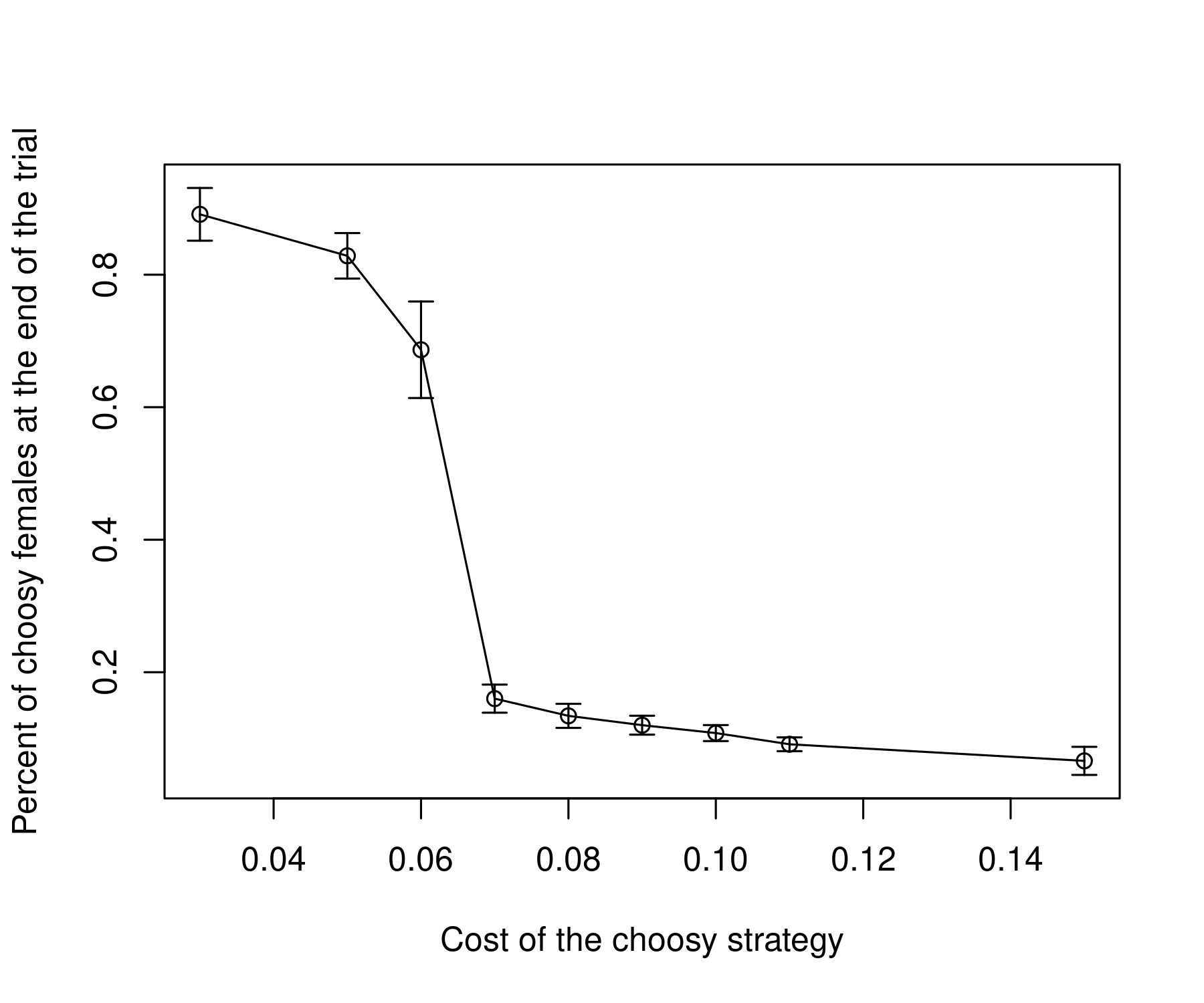}
    \end{subfigure}~
    \begin{subfigure}[b]{0.3\textwidth}
\caption{\tiny Global beacon, $N=5$.}    \label{fig:plotGBLBS5}
\centering
\includegraphics[width=1\textwidth]{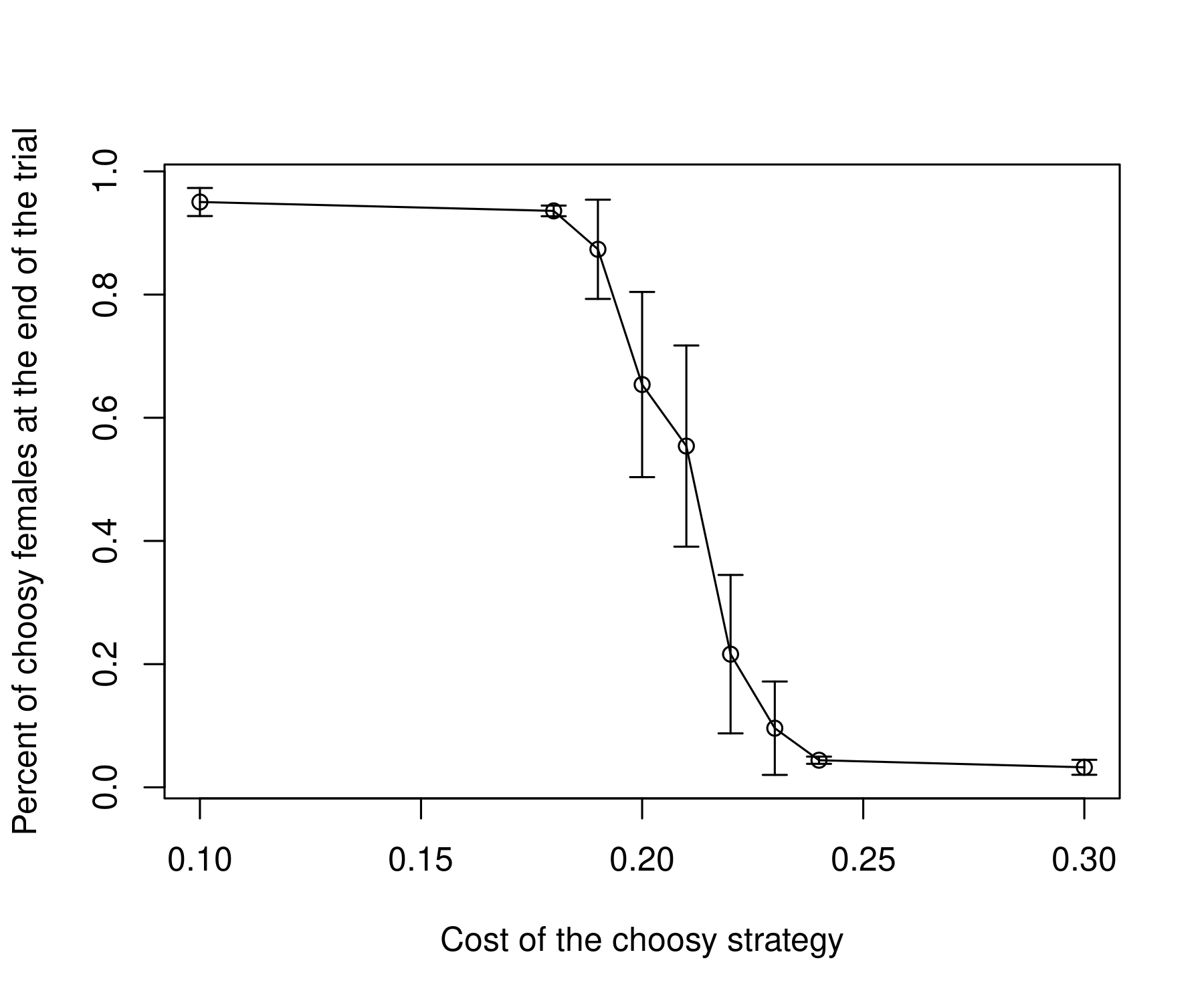}
    \end{subfigure}
    \begin{subfigure}[b]{0.3\textwidth}
\caption{\tiny Global beacon, $N=100$.}    \label{fig:plotGBLBS100}
\centering
\includegraphics[width=1\textwidth]{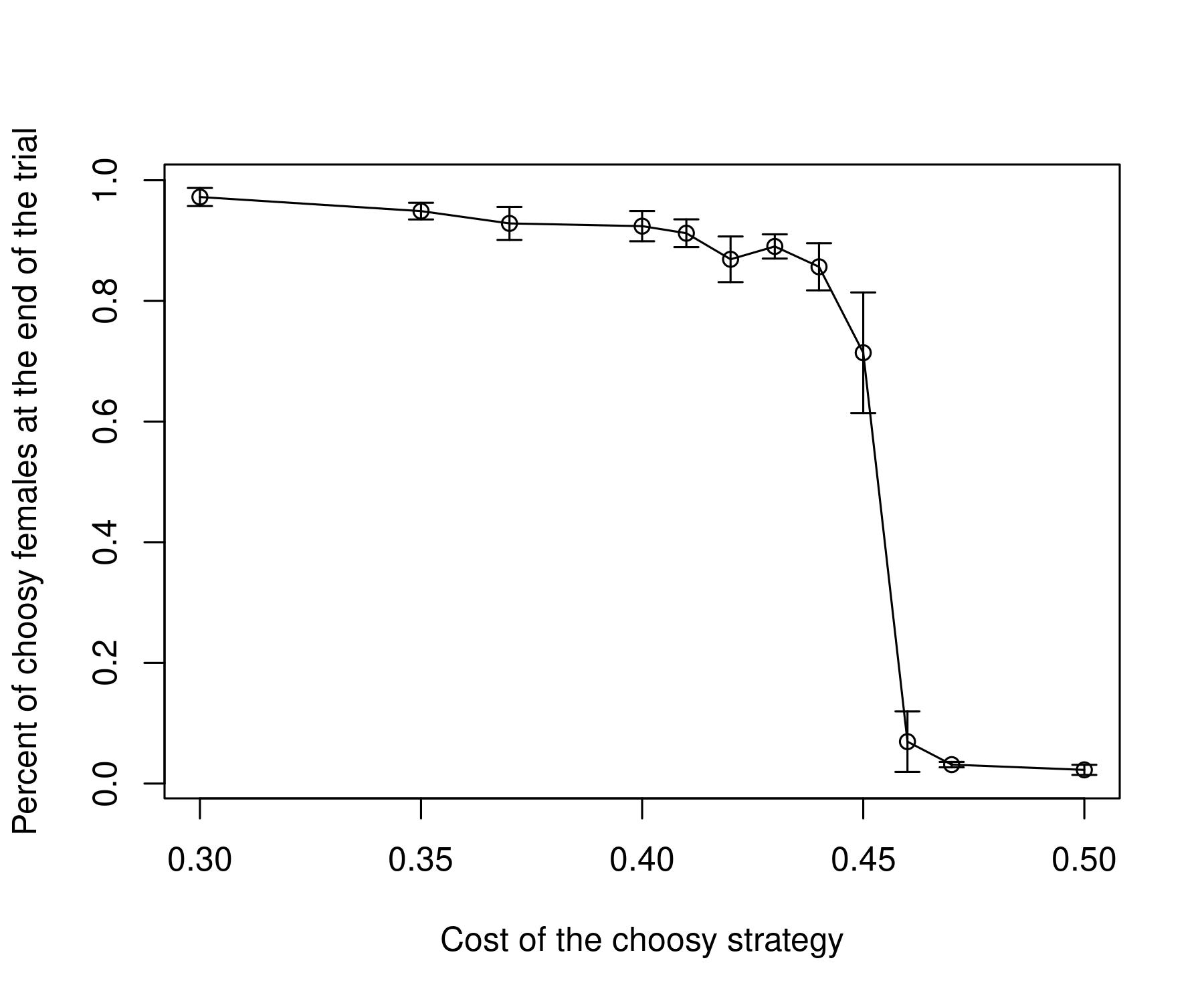}
    \end{subfigure}
\bigskip\vskip5mm
\centering
    \begin{subfigure}[b]{0.3\textwidth}
  \centering
  \caption{\tiny Female-estimated beacon, $N=2$.}
      \label{fig:plotFLBBS2}
\centering
\includegraphics[width=1\textwidth]{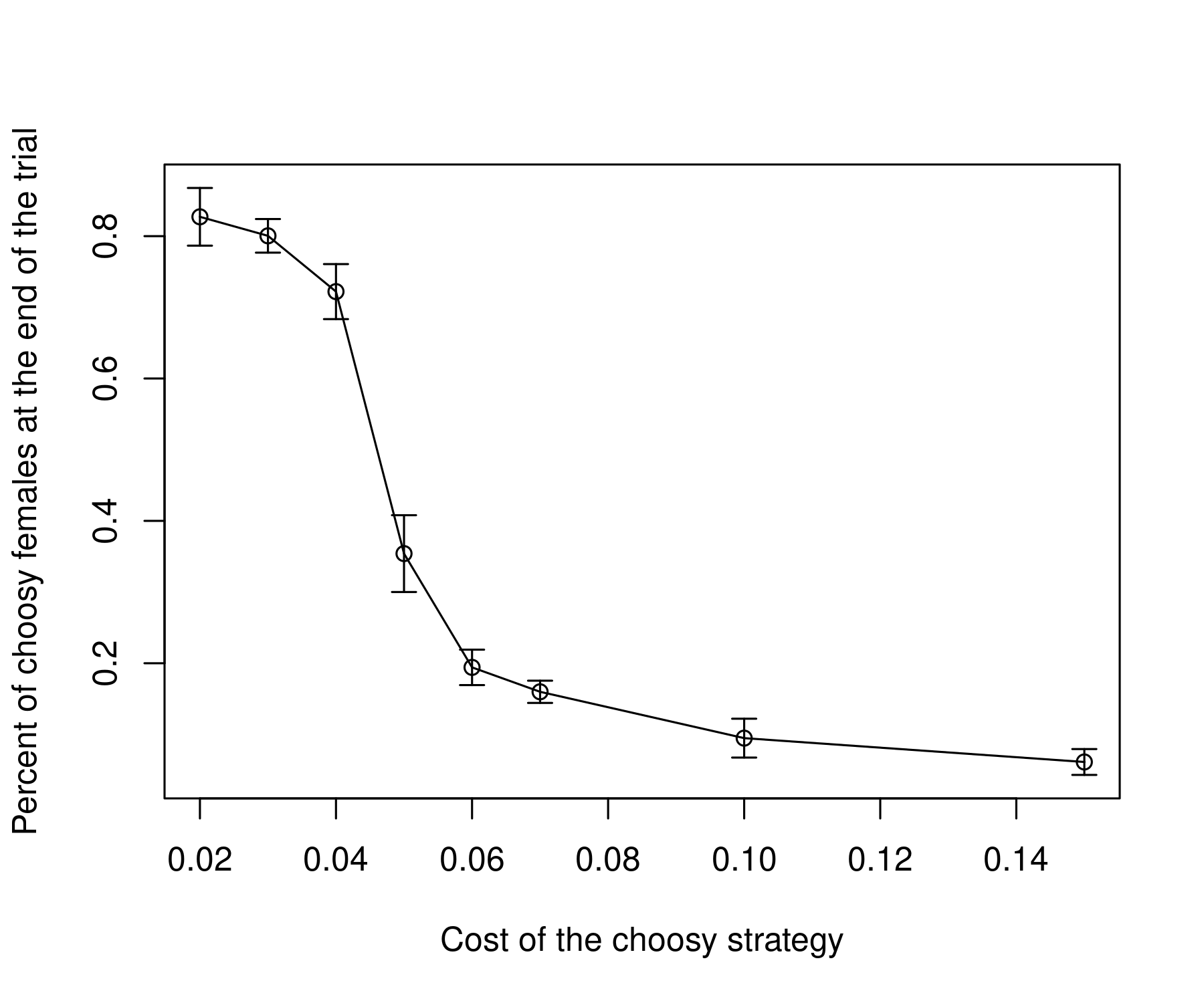}
    \end{subfigure}~
    \begin{subfigure}[b]{0.3\textwidth}
\caption{\tiny Female-estimated beacon, $N=5$.}    \label{fig:plotFLBBS5}
\centering
\includegraphics[width=1\textwidth]{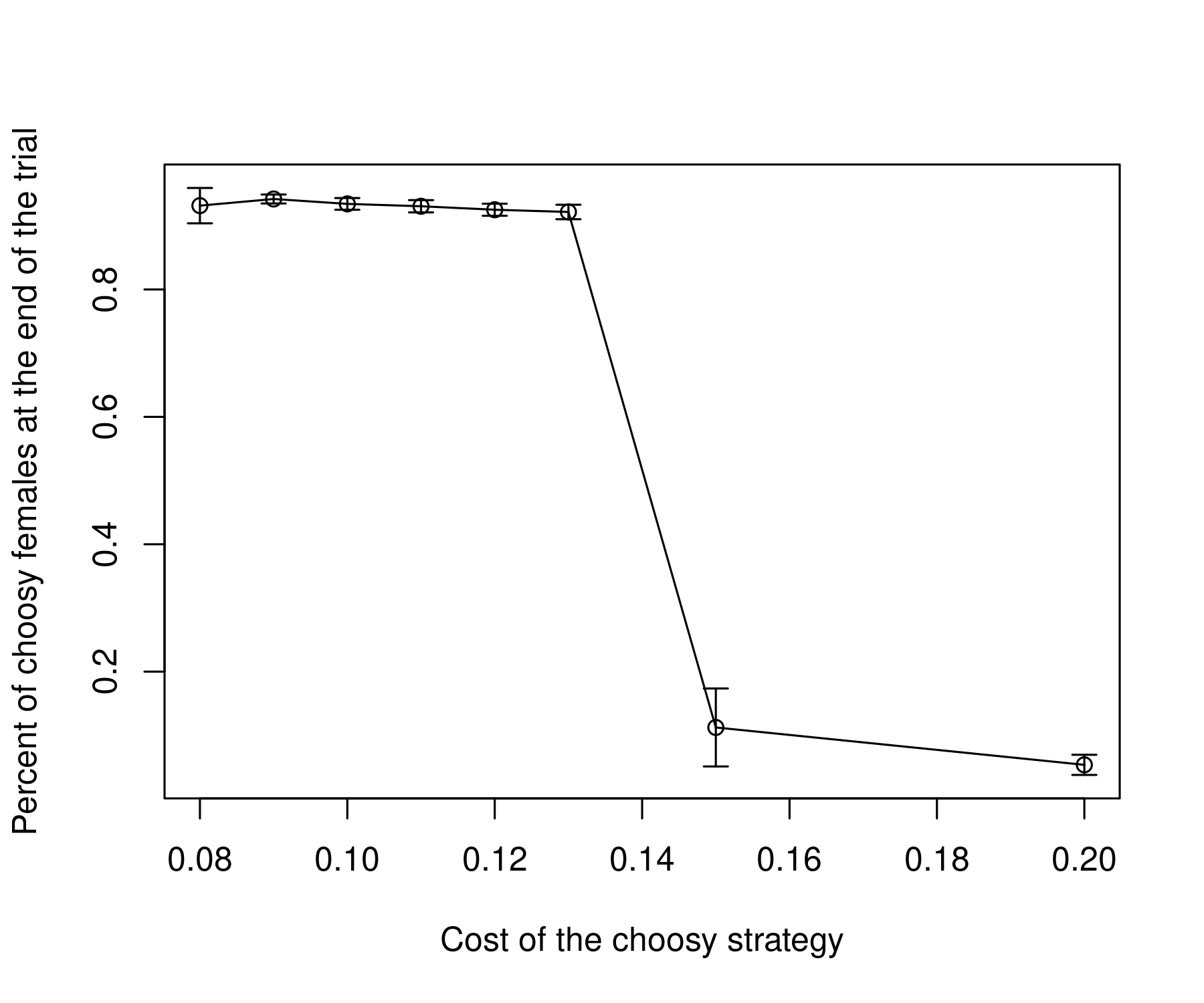}
    \end{subfigure}
    \begin{subfigure}[b]{0.3\textwidth}
\caption{\tiny Female-estimated beacon, $N=100$.}    \label{fig:plotFLBBS100}
\centering
\includegraphics[width=1\textwidth]{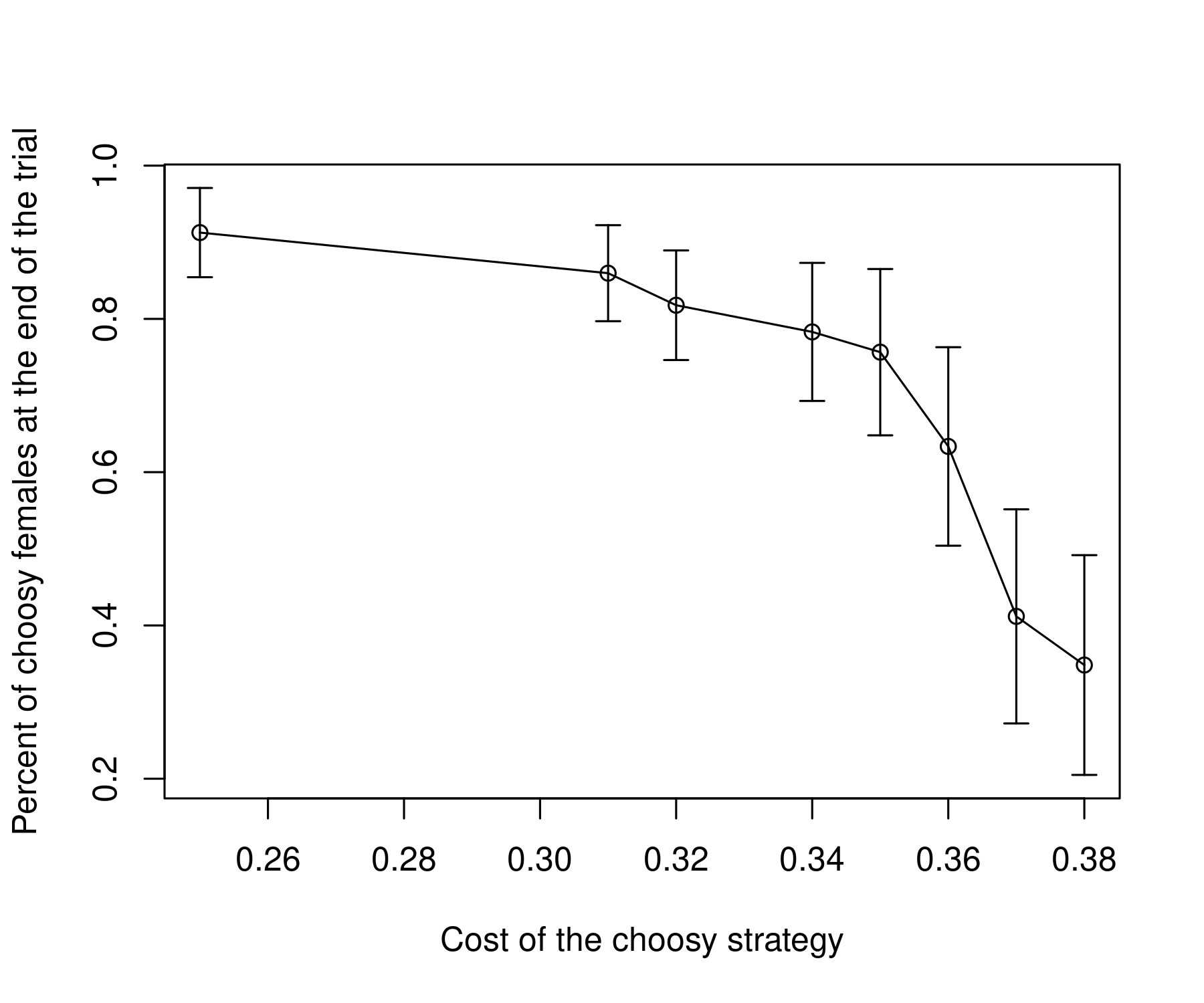}
    \end{subfigure}
\end{figure}

Figure~\ref{fig:plotsBS} shows the viability of the beacon strategy with global beacon with different values of $N$. Each plot shows the percentage of choosy females 
at the end of the $T=5000$ trials. 
The  table in Figure~\ref{tN} gives the estimated values of the  $0.5$-critical viability for different values of $N$ %
 both for the the global beacon and 
for the female-estimaed beacon strategy; the estimated values in the table are with  0.01~confidence interval~over~50~replica~runs.

\begin{figure}[h]
\caption{\small Viability of the beacon strategy with female-estimated beacon with various values of the parameter $B$: 20 (red), 50 (green), 100 (yellow), 250 (blue) and $\infty$ (black). The rest of the parameters are as before:
    $T= 5000$, $N=20$,  $m=m_s=0.01$, $S=2000$.  Confidence intervals are at 1\% with either 10 or 50 replica runs.}
\label{fig:plotFLBBBS}
\centering
\includegraphics[width=0.8\textwidth]{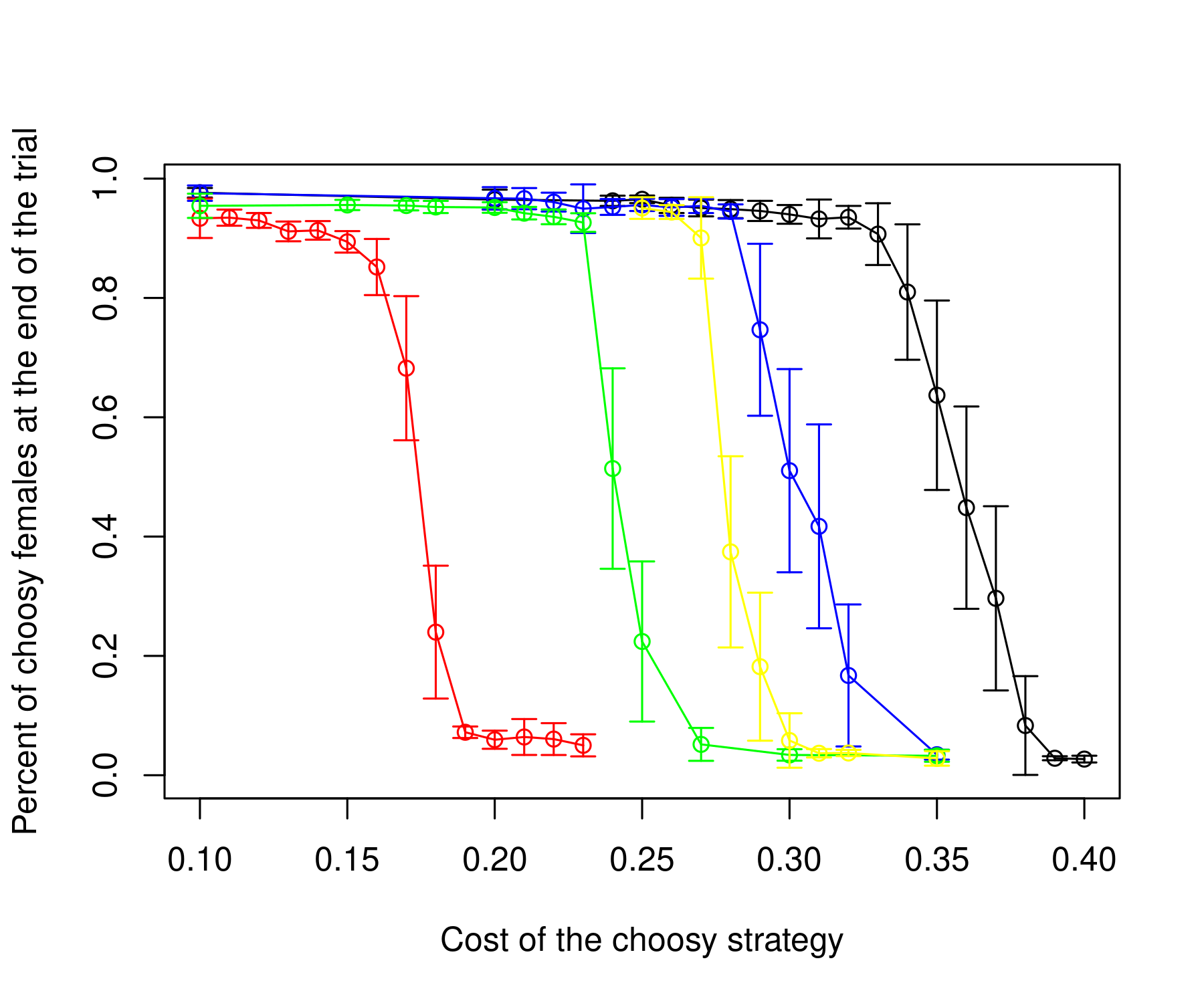}
\end{figure}

Let us next look at the role of the other batch size, $B$, used in the female-estimated beacon strategy in order to construct estimates of the beacon.
We can see that this parameter affects the performance in a predictable way~--- the bigger the batch size, the better, but its influence is not as large.
It is also interesting to consider the female-estimated beacon with the perfect estimation, that is, where each female can observe the whole population in order to estimate the beacon (but not to select a mating partner from), thus getting another version of a ``global'' beacon, but this time based on the average of the ornament beacon of all the males in the population; let us call this case $B=\infty$. On Figire~\ref{fig:plotFLBBBS}, the viability of the beacon strategy is plotted with various values of the parameter $B$, from 2 to $\infty$; recall that the value $B=100$ was used in the experiments in the previous section.

\subsubsection{Mutation rates and the number of attributes}
\begin{figure}[h]
\caption{\small Viability of the beacon strategy with global beacon,  with different mutation rates $m$: 0.01 (black), 0.001(blue) and 0.0001 (green). 
The beacon changes every generation 1 locus at a time. The rest of the parameters are as before:
    $T= 5000$, $N=20$,  $m_s=0.01$, $S=2000$.  Confidence intervals are at 1\% with either 10 or 50 replica runs.}
\label{fig:plotGLBMUT}
\centering
\includegraphics[width=0.8\textwidth]{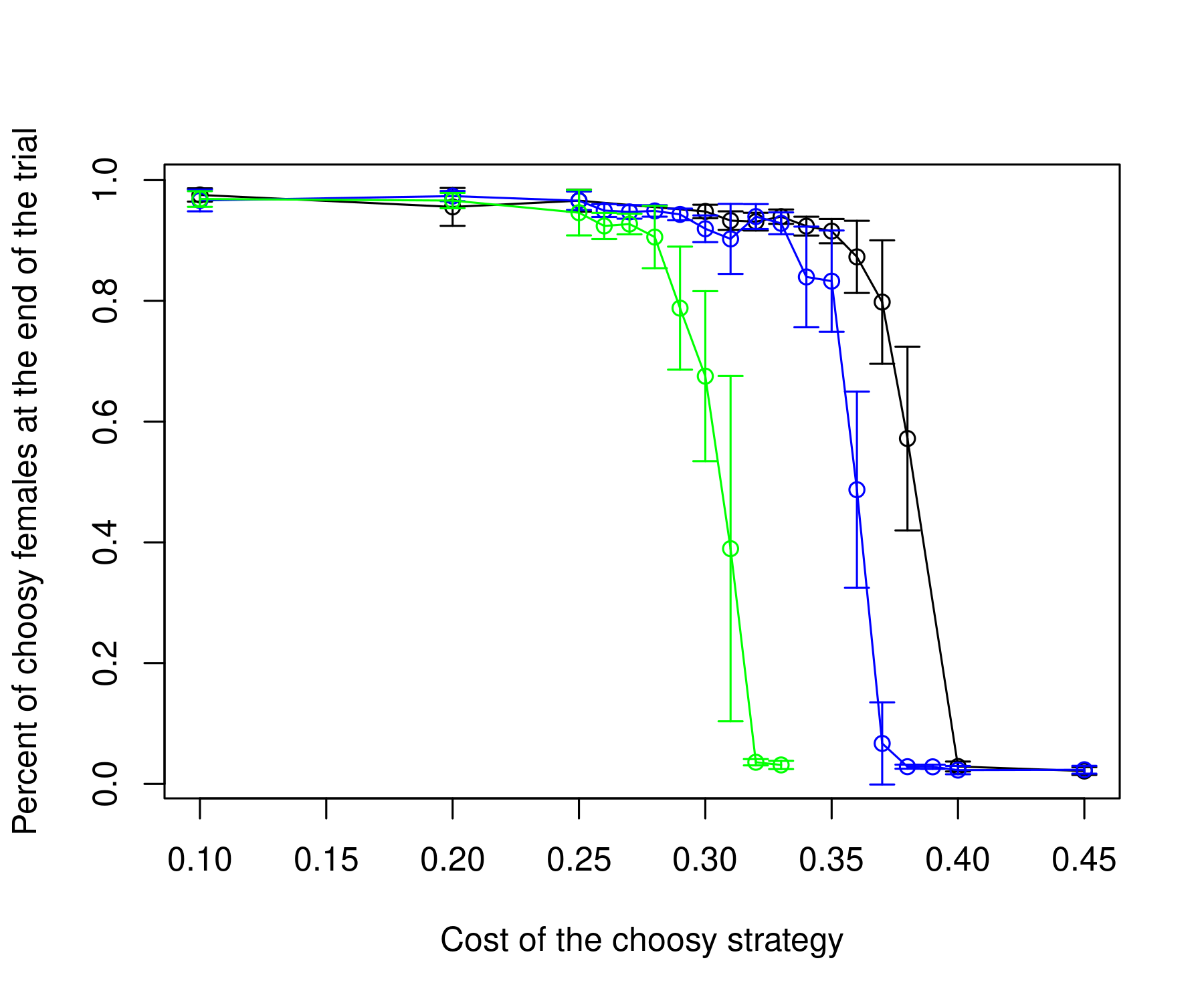}
\end{figure}

Lower mutation rates result in lower viability of the beacon strategies.  Figure~\ref{fig:plotGLBMUT}  presents the viability plots for mutation rates $m=0.001, 0.0001$ and $0.01$ as before for comparison.  These lower mutation rates are more realistic. However, as mentioned in the introduction, higher mutation rates can be achieved by increasing the number of attributes and combining them. 

\begin{figure}[h]
\caption{\small Viability of the beacon strategy with global beacon,  with different number of attributes $A$ in the male beacon: 
20 (red) and 100 (blue), with mutation rate $m=0.01$. The black line corresponds to $A=1000$ with $m=0.001$. 
The beacon changes completely every 50 generations. The rest of the parameters are as before:
    $T= 5000$, $N=20$,  $S=2000$.  Confidence intervals are at 1\% with either 10 or 50 replica runs.}
\label{fig:plotNumAttr}
\centering
\includegraphics[width=0.8\textwidth]{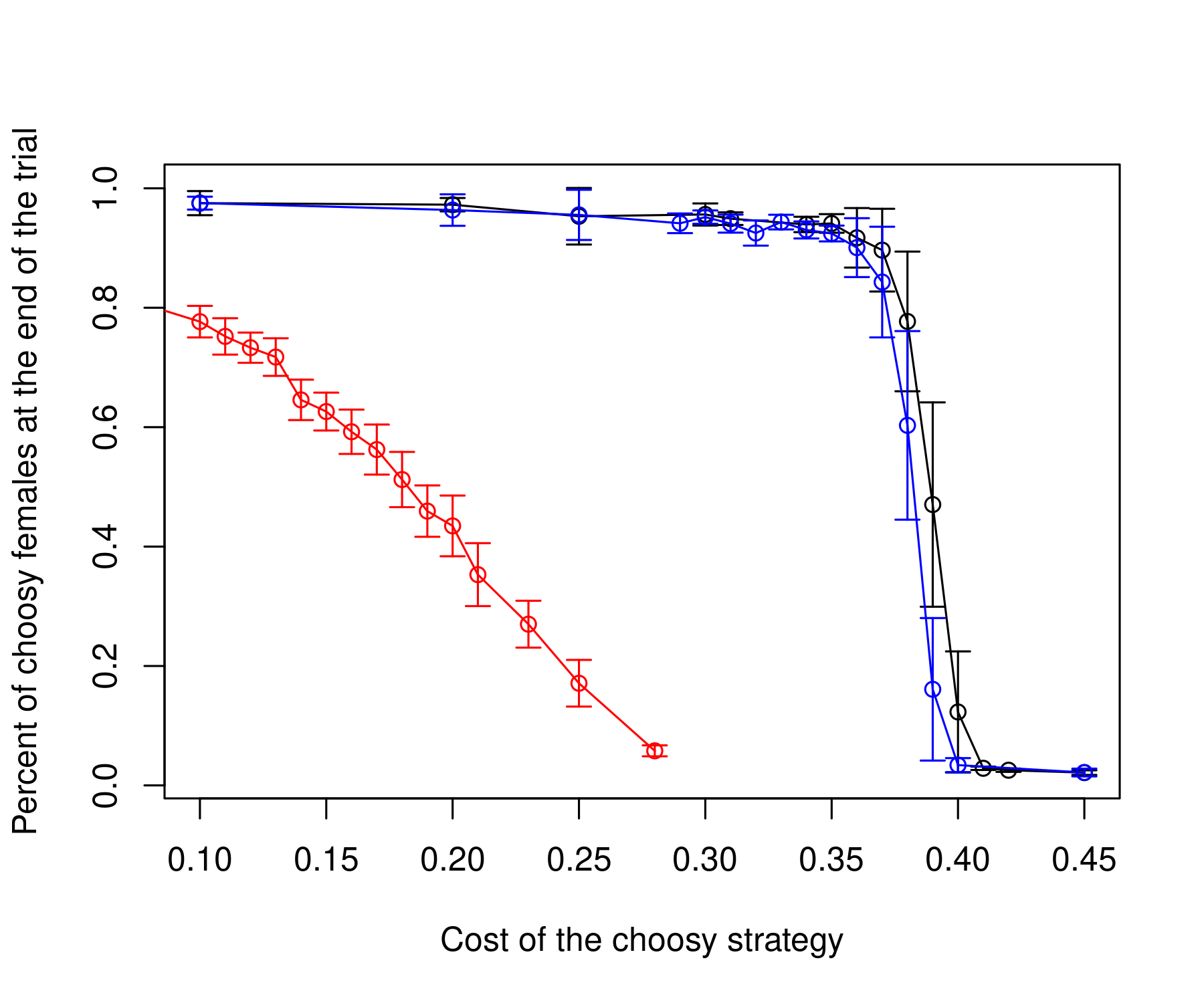}
\end{figure}

Indeed, as seen on Figure~\ref{fig:plotNumAttr}, decreasing the number of attributes decreases viability. 
On this plot, we also show the viability of the beacon strategy with a higher number of attributes ($A=1000$) and smaller mutation rate ($m=0.001$) which is more realistic for this number of attributes; it can be seen that this leads to a slightly increased viability, as compared to the values $A=100$ and $m=0.001$.

For female-estimated beacon, lowering the mutation rates results in a quick deterioration of viability, as the resulting beacon does not have enough randomness. The results are predictable and therefore omitted.

\subsubsection{Parameters of the global beacon  and another look at the biased-mutation model}
We tested various values of the parameters of the global beacon: gradual  changes, that is, a change every  generation with $1, 2, 3, 5$ or $10$ loci at a time;
and changing completely every $10, 20, 50$ or $100$ generations. The parameters affect the viability of the strategy but the changes are small, so we do not report them.

For the beacon that changes completely every $K$ generations, observe that when $K$ goes to infinity, the resulting strategy comes back to a version of the biased mutation strategy. Indeed, since each locus of the ornament may mutate every time but the comparison (beacon) ornament never mutates, once the beacon-like ornament spreads over the population, the mutation becomes  biased against the ornament, in the sense that each mutation makes the ornament more different from the beacon with a probability higher than that of making it closer to the beacon. 

\begin{figure}[h]
\caption{\small  Percentage of females following the beacon strategy with slowly changing beacon: the beacon  changes completely every 300 generations. One cercle per generation, over $T=1500$ generations.
Cost of the choosy strategy is $c=0.25$.  The rest of the parameters are as before:
$m=0.01$,  $N=20$,  $S=2000$. The initial population is random}
\label{fig:runnoran}
\centering
\includegraphics[width=0.8\textwidth]{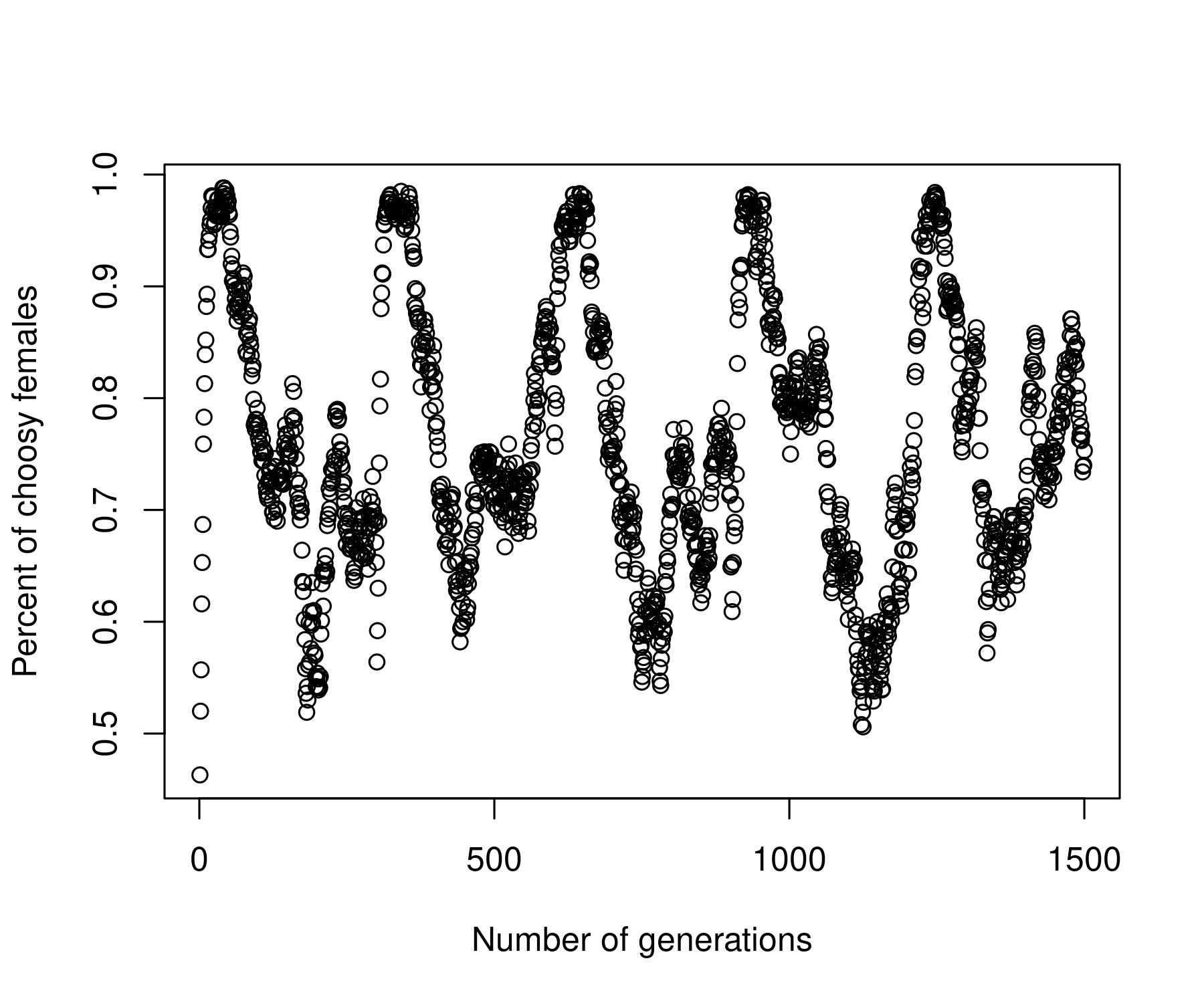}
\end{figure}

It is, therefore, interesting to look at the viability of the beacon strategy with a beacon that changes rarely, as it presents a middle ground between the 
biased mutation model of the  Fisher runaway process and the fashion-led beacon strategy studied. Instead of presenting viability plots, we show, on Figure~\ref{fig:runnoran}, the average number of choosy females during a single trial run, with the beacon changing completely every 300 generations. Over 1500 generations, we can clearly see the 5 peaks corresponding to the changes in the beacon (shortly after generations 0, 300, 600, 900 and  1200): a shot of new randomness gives a boost to the choosy strategy. Note that, between the peaks, while the choosy part of the population decreases, it does not die off quickly despite the relatively high cost of 0.25, as we are in the biased-mutation mode here, which is a viable strategy in itself.

\section{Discussion and future work}
We have shown that correlated female choice strategies can effectively channel  randomness from the environment into genetic diversity, providing the needed fuel for the runaway process to keep running, and thereby maintain the choice in the population. This gives new answers to some long-standing questions in evolutionary biology, including the lek paradox, but it also opens up various avenues for further modeling and research. Some of these are discussed in this section.

We start with some direct generalizations that can be made to the model in order to make it more realistic or amenable to theoretical analysis. 
We then continue with implications into related topics in evolutionary biology, including the driving forces for the evolution of intelligence, more complex 
mate choice strategies such as mate choice copying and other related questions.

\section{Direct extensions and  generalizations}
The model used in this work  is perhaps the simplest one for the phenomenon in question (at least, in the framework of individual-based simulations); in this quality, it is suitable for answering qualitatively the questions at hand.  However, a few generalizations would make it more realistic, and are therefore worth exploring in future research. It would also be interesting to see quantitatively how much each generalization affects the viability of the beacon strategies considered. 

The first generalization that comes to mind is diploidy and  genetic recombination of ornaments. Since these genetic mechanisms allow for more stochasticity, it can be conjectured that they would increase the viability of choosiness even further. 

Another generalization concerns the cost of choosiness and the sampling methods used by choosy females. Following classical models of the Fisher runaway process \citep{lande1981models,kirkpatrick1982sexual,pomiankowski1991evolution}, here we used fixed  costs  of choice; moreover, each female 
is given a batch of a fixed size to choose from. It would perhaps be more realistic to let the cost depend on the batch size and allow the females to choose when to stop sampling. \cite{andersson1994sexual} has suggested that preferences  may have a cost that is inversely
related to the frequency of the preferred type of male.  \cite{kokko2015mate} and subsequent works (e.g., \citealp{henshaw2022evolution}) have incorporated dynamic sampling costs into models of the runaway process, and their results show that this generalization  is significant. 
In our model, every male is potentially unique, so that the decision of each female when to stop  sampling is not as simple as ``stop when the preferred type is found,''  and would necessarily involve extra parameters.
Here we have also used exclusively the best-of-$N$ sampling model, whereas there are various other possibilities both in theory \citep{janetos1980strategies} 
and, of course, in nature \citep{rosenthal2017mate}.
Nevertheless, since the batch size $N$ affects greatly the viability of the strategy, it appears important to explore the effects of various sampling methods and 
 the corresponding cost structures in our model.

\cite{kuijper2012guide} survey four most widely used  approaches to modeling sexual selection and the Fisher runaway process, of which we have chosen the most complex: individual-based simulation. The other three (population genetics, quantitative genetics and invasion analysis) are mathematical approaches  that attempt to fully describe the dynamics of the process at the cost of making some simplifying assumptions. While we did not find these models directly applicable to our scenario, invidual-based simulation still has some important disadvantages that only mathematical modeling can remedy. Specifically, the results of the simulations depend on the value of the  parameters, which include the initial population, and this space is impossible to explore exhaustively. Rather than attempting to model the process precisely, we would like to call for a different, qualitative, approach, that would help to answer questions of the following kind. Consider the exact same model used in the individual-based simulation. Find upper bounds on the cost sustainable by the beacon strategies: can, for example, the cost of 50\% be sustained by these strategies with some values of the parameters? (Note that a trivial upper bound of 1 can be established mathematically.) Does a strategy A   dominate a strategy B, in the sense that
for every set of  initial conditions the critical viability of strategy A is higher than that of strategy B?  and so on. To study the model mathematically, one can note that if the beacon changes every generation then it is a  finite-state Markov chain (the state space is finite but huge). The theory of Markov chains (e.g., \citealp{hernandez:03}) provides the necessary apparatus to address qualitative questions of the kind mentioned. 
The advantage of such an approach would be that the model studied mathematically is the same as the model studied empirically in simulations; this would come at the cost of giving up on full descriptions of the dynamics of the model.
We leave this task, which appears highly non-trivial, as an avenue for further research.

\subsection{Complexity of choice, prediction and intelligence}
Mate choice emerges from these results as  a rather complex task. Indeed, it requires many observations of potential mates~--- the more the better, and these observations are based on a large set of stimuli~--- again, the more stimuli are used, the better. As can be observed in nature, these stimuli can be spread across a variety of modalities: visual, acoustic, etc. (e.g., \citealp{hegyi2022functional}); that this should be so does not follow from the results of the model, but this can be simply a way to increase the number of attributes to select from.  Furthermore, observations are required to construct the beacon for the beacon strategies; what are these observations we do not know, but at least if they are constructed from the population itself, then, once again, sampling is needed, and the larger the sample the better.  The function that maps the beacon into male sexual features need not necessarily be complex (it may be some simple scrambling, like the one mentioned in the introduction), but the data processing task resulting from the sampling described is potentially rather demanding.

However, the mate-choice problem is perhaps yet a lot more complex than that,  possibly pushing the cognitive abilities of the species to their limit; it may even be a force driving the evolution of intelligence~--- albeit limited  a one whose application is limited to females (or, more, generally, to the choosy sex), as we shall presently argue. 
Recall that the beacon strategies rely on an external source of randomness. Randomness is rarely really random in nature; much of the stochasticity in the environment is at least partially predictable. Suppose that some females are able to predict the next change of the beauty beacon. Then they would be clearly advantaged with respect to the rest, as they would be able to select an optimal  match for both the current and the next beauty target. However, if sufficiently many females are able to make the prediction, then the next target becomes the current target, and the prediction problem shifts to the subsequent one. Then the choosy females either deplete the source of randomness, in which case the beacon strategy ceases to become advantageous and they need to look for another source of randomness; or else, 
a prediction race opens up, eliminating all that the females of the population are able to predict well from the beauty beacon, and pushing what remains to the limit of their abilities. As mate choice is a never ending competition,  it may well result   being one of the most cognitively demanding problems that an individual can possibly face.

From this argument, it may be hypothesized that females should be more intelligent than males in at least some lek species. However,  male intelligence may arise as well as a by-product, and  males might also benefit  from it in other ways, some of which are discussed below in Section~\ref{s:malechoicetc}.
To our knowledge, cognitive difficulty of mate choice has not been studied explicitly, although there is a large literature on mate choice copying which is considered a case of (social) learning; this is discussed in Section~\ref{s:copy}.

\subsection{Mate choice copy and speciation}\label{s:copy}
Mate choice copying is said to occur when a female is more likely to mate with a previously mated male \citep{bennett1998mate}, and is a part of a more general phenomenon
of non-independent mate choice \citep{vakirtzis2011mate}. It has been observed in a variety of species \citep{davies2020meta}, from insects to humans \citep{mery2009public,eva2006all}, and studied both theoretically and empirically \citep{lill1974sexual,pruett1992independent,dugatkin1992sexual,kirkpatrick1994sexual}. 
Cost-avoidance  and improving estimation accuracy are the typical explanations proposed for this phenomenon. 
It has been argued \citep{sapage2021social} that mate choice copying may be advantageous in rapidly changing environments, where 
females might need updated information regarding better adapted or more popular mates.

While we do not attempt to model mate choice copying here, our results show the importance of estimating the attractiveness of potential males and the necessity of relatively large samples. Clearly, bootstrapping the choice of others may be helpful here. It can be conjectured that mate choice copying can significantly lower the batch size $N$ required to sustain a given cost $c$.  

Perhaps more importantly, the cognitive difficulty of making the choice discussed above introduces the possibility that some females may be unable to make the choice purely by themselves, or, else, be less apt at it than others. Giving the constantly changing nature of the target, mate choice copying may enable the co-existence of choosers apt at different versions of the task, or different parts of it. Mate choice copying thus may not be simple copying but rather learning from others; indeed, it is often referred to as a case of social learning  \citep{witte2015mate,davies2020meta,sapage2021social}, and sometimes includes generalization, whereby the observed preference is generalized to potential mates with the same features \citep{vakirtzis2011mate,white2000culture,fowler2015complexities}.
A closely related concept that appears applicable to the problem of choice in  our model is that of distributed social learning. The latter is said to occur \citep{reznikova2022flexibility} when only a few members of a population are able to solve the task without looking at others (presumably, they have inherited the complete  behavior genetically), whereas the rest are only able to complete some parts of it and  learn the rest (complete the pattern) by observation. Thus, the genetic information needed for this behavior is distributed over the population. This phenomenon has been demonstrated in ants and rodents in the context of non-obligatory hunting behavior  \citep{reznikova2008ant,reznikova2022flexibility}. In the context of mating behavior, it has been shown in fruit flies
that males with a missing gene responsible for essential details of mating behavior can learn these missing elements  when kept in a group \citep{danchin2018cultural}.

Since the mate choice problem in our model is both complex and dynamic, simple mate choice copy as well as more general forms such as distributed social learning appear highly relevant, and deserve further modeling research. It is worth nothing that, despite the large literature on mate choice copying, the question of how difficult the problem is for the choosers, that is, its intellectual aspect, has remained largely unexplored so far.

Turning to speciation, it has been noted already by \cite{darwin1871descent}  and the early researchers on evolution (cf.\ \citealp{lande1981models} and references) that closely related species often differ the most in the characteristics of adult males. However, modeling attempts conclude that sympatric (that is, within the same geographical area) speciation by sexual selection is not likely \citep{arnegard2004sympatric}  even taking into account mate choice copying \citep{mery2009public}. More broadly, while sympatric speciation is a contentious topic, it is generally considered to be possible even if it is unclear how common it is \citep{bolnick2007sympatric}.

While we do not attempt to model speciation here (leaving this topic for further research), we can make the following two observations related to the argument. 
First, given the difficulty of the task of female choice described above and the uncertainty that arises from it, the choice function applied by females appears fragile and prone to both error and mutation; thus, sympatric speciation by sexual selection appears to be worth reconsidering in this light.
Second, given that the choice function takes into account external information, it may depend on the environment, which may lead to allopatric speciation as a direct consequence.

\subsection{Male adaptations, male choice and other generalizations}\label{s:malechoicetc}
As perhaps in all models of the runaway process, in the one considered here males have no way to influence their fate once their genotype has been determined.
 This is not always so in nature, as, even in lek species, males may change their courtship  behaviour during their lifetime (for example, \citealp{dukas2005experience,kahn2013strategic}). Females can also change their preferences, 
 which is of course not addressed in our discrete-generation model. Adaptive behaviour by both males and females provide interesting directions for generalizations. 
 One can hazard a guess that the result would be that the choice problem becomes even more complex and even more randomness is needed. 
 
Another promising generalization is mutual choice. In many species, especially in those where males contribute to raring the young, both males and females are choosy. 
Whether, or to which extent, this leads to a runaway process on both sexes is not obvious. Furthermore, the interplay between male and female strategies may be non-trivial here, and may include the appearance of such   strategies as playing-hard-to-get (e.g., \citealp{jonason2013playing})  which may become a part  of the runaway process \citep{ryabko2015evolutionary}. These are all interesting topics for further modeling. 

  \bibliographystyle{apalike}
  \bibliography{ff}

\end{document}